\documentclass[aps,prl,superscriptaddress,showpacs,floats,twocolumn,floatfix]{revtex4-1}

\usepackage{color}
\usepackage{amsmath,amssymb,stmaryrd,bm}
\usepackage{graphicx}
\usepackage{hyperref}
\usepackage{setspace}
\usepackage{accents}
\usepackage{marvosym}

\usepackage{tabularx}

\usepackage[normalem]{ulem}
\newcommand\redout{\bgroup\markoverwith{\textcolor{red}{\rule[.5ex]{2pt}{1.4pt}}}\ULon}

\graphicspath{{fig/}}
\newcommand{\Eff}{{\cal N}}

\begin{document}

\addtolength{\topmargin}{10pt}

\title{Efficacy of self-phoretic colloids and microswimmers}

\author{Amir Nourhani}
\email{nourhani@psu.edu}
\author{Paul E. Lammert}
\email{lammert@psu.edu}
\affiliation{Center for Nanoscale Science,The Pennsylvania State University, University Park, PA 16802}
\affiliation{Department of Physics, The Pennsylvania State University, University Park, PA 16802} 


\begin{abstract}

Within a unified formulation, encompassing self-electrophoresis, 
self-diffusiophoresis, and self-thermophoresis, we provide a simple integral kernel 
transforming the relevant surface flux to particle velocity for any spheroid with 
axisymmetric surface activity and uniform phoretic mobility.
We define efficacy, a dimensionless efficiency-like quantity expressing the
speed resulting from unit absolute flux density on the surface, which allows a
meaningful comparison of the performance of different motor designs.
For bipartite designs with piecewise uniform flux over complementary surface regions,
the efficacy is mapped out over the entire range of geometry (discotic through sphere to 
rod-like) and of bipartitioning, and intermediate aspect ratios that maximize efficacy
are identified. Comparison is made to experimental data from the literature.

\end{abstract}
\maketitle 

The challenge of powering motion at the (sub)micro-scale has motivated the 
development  of a variety of abiotic micromotors as building 
blocks of micromachines over the past decade.~\cite{Ebbens:2010p86,JWangBook2013,Paxton:2004p183,C5CC00565E} 
Artificial self-phoretic colloids,
harvesting energy from the environment and transducing it to motion via active surfaces, 
offer a unique solution to  
this challenge~\cite{Popescu2010EPJ,Golestanian+-07,Nourhani2015PRE062303, Nourhani+-15-PoF-spheroid, Nourhani+-15-PoF,EbbensPRERp020401,SchnitzerYariv2105}.
The performance of a self-phoretic particle
is determined by its shape and distribution of surface activity,
whether the operative mechanism is
self-diffusiophoresis~\cite{Popescu2010EPJ},
self-electrophoresis~\cite{Nourhani2015PRE062303, Nourhani+-15-PoF-spheroid, Nourhani+-15-PoF,Yariv:2011p74,Sabass:JCP:2012}, or self-thermophoresis~\cite{Jiang2010PRL268302,Golestanian2012PRL038303}
While quantitative analysis of these factors are essential for designing 
fast and efficient motors, studies are mainly limited to spheres or long
thin rods and neglect intermediate shapes and disks,
though sphere dimers~\cite{Reigh+Kapral-15} 
have also received attention. 

\begin{figure}[b]
\begin{center}
\includegraphics[width=0.47\textwidth]{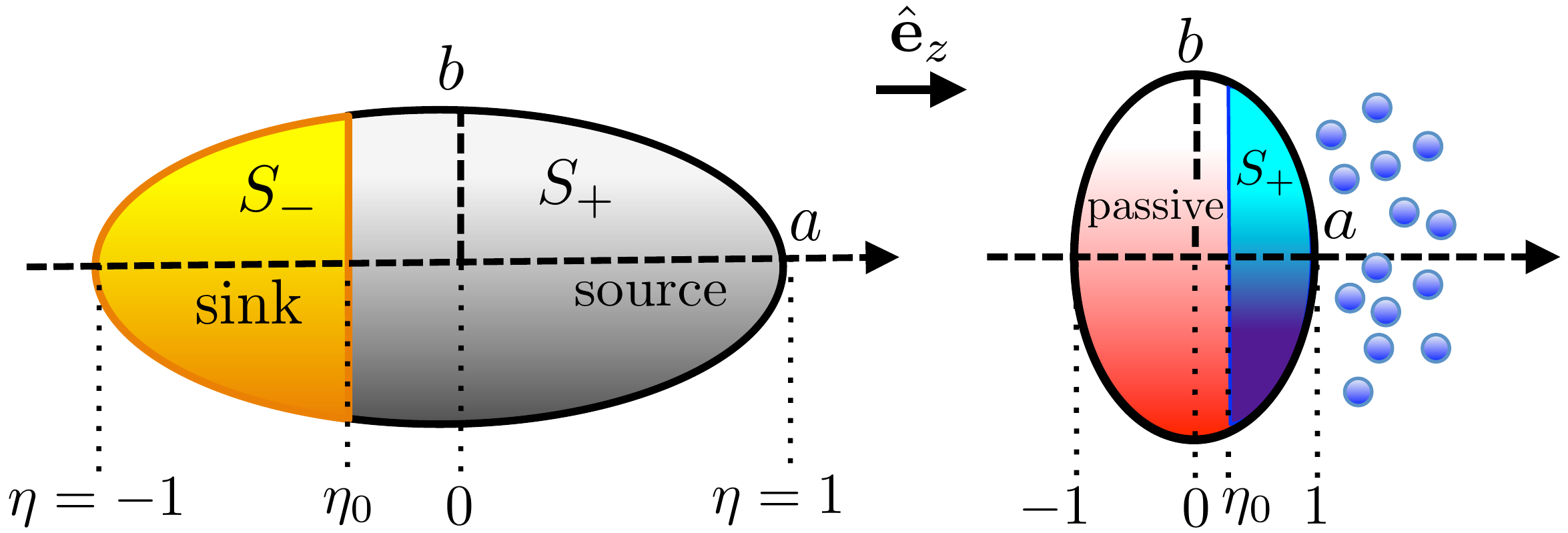}
\end{center}
\vspace{-15pt}
\caption{ 
(color online).
Standard bipartite geometry on the spheroid. 
The border between the two regions corresponds to scaled $z$ coordinate $\eta = \eta_0$.
($\eta$ is equivalent to $\cos\theta$ for a sphere.)
(left) In the source/sink case, both sides on the surface are active
while (right) in the source/inert case only one side is active.
\label{fig:SpheroidGeo}
}
\end{figure}

\begin{figure}[t]
\begin{center}
\includegraphics[width=0.38\textwidth]{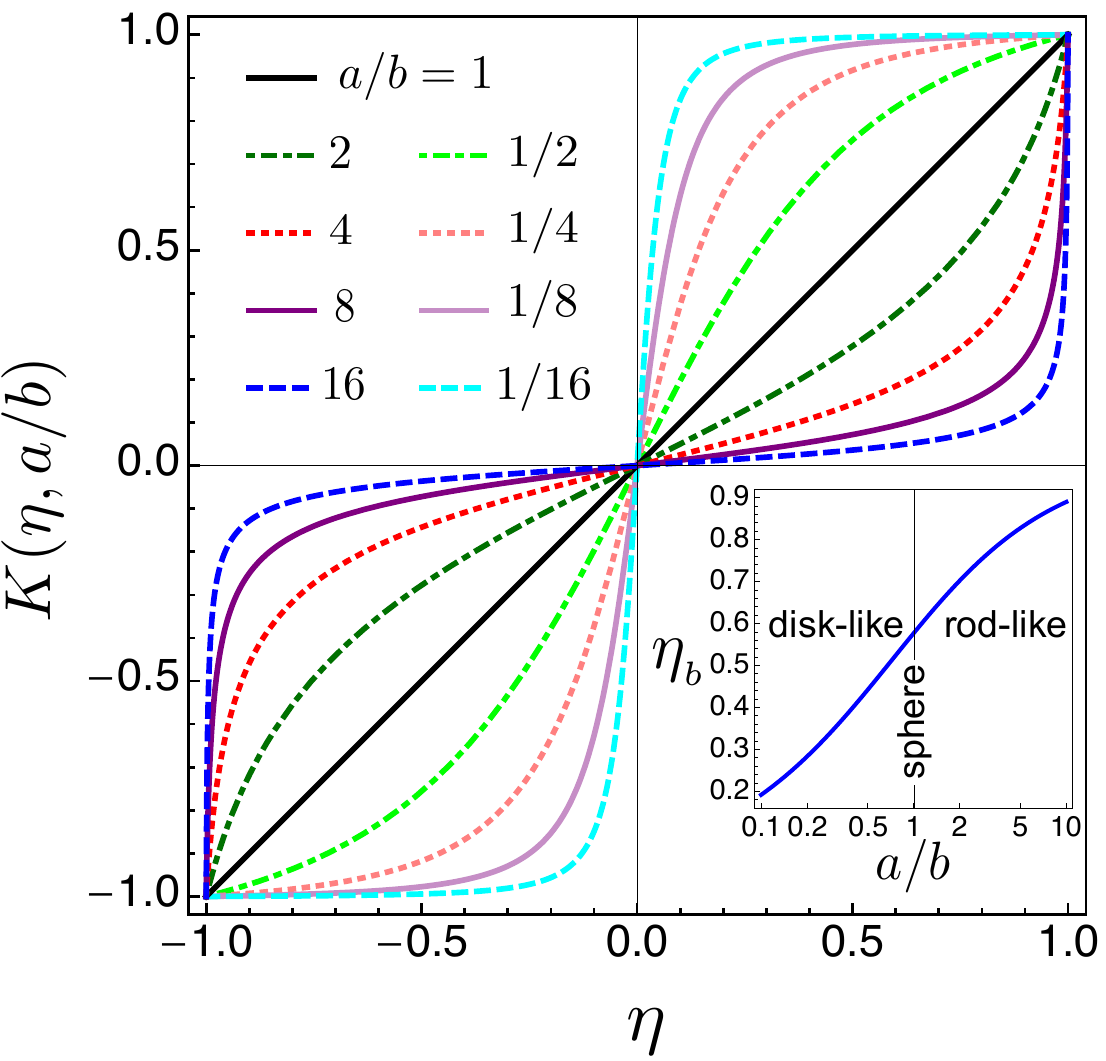}
\end{center}
\vspace{-15pt}
\caption{ 
(color online).
The spheroid velocity kernel, Eq. (\ref{eq:kernelspheroid}), 
for a range of aspect ratios. 
$K(\eta;\ell^{-1})$ is just $K(\eta;\ell)$
reflected across the diagonal $\ell\equiv a/b=1$.
For a sphere ($a/b= 1$) the kernel
is simply linear. On the prolate side ($a/b > 1$), a zone of suppressed effectiveness
moves outward from the equator with increasing aspect ratio, and on the
oblate side ($a/b < 1$) a zone of enhanced effectiveness moves inward from the poles
as aspect ratio is decreased. 
Inset: 
The ``belly'' $\eta_b$ is the point at which $dK/d\eta = 1$ and
provides a quantitative expression of the division into effective 
and ineffective regions. The plot shows $\eta_{_b}$ as a function 
of aspect ratio. 
\label{fig:keta}
}
\end{figure}

In this Letter we seek to elucidate the determination of self-phoretic particle 
performance by both overall shape and surface distribution of activity,
under common approximations of uniform phoretic mobility, thin interaction 
layer and linearity.
The practical benefit is a rational approach to higher ``efficacy'' ---
greater speed for the same energy (fuel) consumption.
Within an approach unifying various self-phoresis mechanisms,
we explore the design space of axisymmetric surface activity
for the entire spheroid family, which smoothly interpolates from disks 
through spheres to needle-like shapes.
The fundamental innovation on which the treatment turns is a simple integral kernel 
[Eqs. (\ref{eq:velocity-K-integral}, \ref{eq:kernelspheroid}) and Fig. \ref{fig:keta}]
quantifying the {\em local\/} effectiveness of surface activity at producing motion.
Applying it to bipartite flux distributions, the efficacy of the full range of
aspect ratio and bipartitioning (i.e., $\eta_0$ in Fig.~\ref{fig:SpheroidGeo})
is mapped out in detail, and intermediate optimum geometries are identified.
Previously obscure trends, such as a non-monotonic dependence 
of efficacy
on aspect ratio are thereby clarified.
Explicit, closed-form expressions for the speed and efficacy are given 
for these designs in Supplementary Information.

 In Anderson's unifying picture~\cite{Anderson-89} for {\em passive} particles,
phoresis is mediated by an {\em externally-imposed} gradient of a field $\gamma$, 
such as concentration of a chemical species (diffusiophoresis),
electric potential (electrophoresis) or temperature (thermophoresis).
To leading order in interaction layer thickness
(assumed small compared to particle size), the tangential gradient 
of $\gamma$ generates a slip velocity 
${\bm v}_{\mbox{\scriptsize slip}} = \mu_{ph} {\bm \nabla}_{\!s}\gamma$ at the outer edge
of the boundary layer.
 The
phoretic mobility $\mu_{ph}$
usually depends quadratically on the interaction layer's length scale~\cite{Popescu2010EPJ,Nourhani2015PRE062303, Nourhani+-15-PoF-spheroid,Yariv:2011p74}.
In the case of a spheroidal particle, the resulting phoretic velocity is 
known~\cite{Fair+Anderson} to be
\begin{equation}
\bm{\mathcal{U}}
 = 
-{1 \over 3{\cal V}} 
\int_{S}\mu_{ph}
( \hat{n}\cdot {\bm r})
\,{\bm \nabla}_{\!s}\gamma
 \, dS, 
\end{equation}
where $\hat{n}$ is the outward-pointing unit vector normal to the
surface, ${\bm r}$ is position relative to the spheroid center,
and ${\cal V}$ is particle volume.
We approximate $\mu_{ph}$ to be 
uniform, as is commonly 
done~ \cite{Popescu2010EPJ,Nourhani2015PRE062303, Nourhani+-15-PoF-spheroid,Yariv:2011p74}.

While in phoresis of passive particles, $\gamma$
is controlled externally, a {\em self-phoretic} particle sustains
the gradient of $\gamma$ itself by generating a heterogeneous surface flux $\Gamma$.
Therefore, 
it may be more useful and perspicuous to relate the particle velocity directly to the 
pattern of surface activity $\Gamma$ directly, rather than indirectly
through $\gamma$~\cite{Golestanian+-07,Popescu2010EPJ,Nourhani2015PRE062303, Nourhani+-15-PoF-spheroid, Nourhani+-15-PoF,EbbensPRERp020401,SchnitzerYariv2105}.
Motion of present-day motors has little effect on chemical kinetics at their surfaces 
and only the resulting flux is needed for our study~\cite{Nourhani2015PRE062303}.
Because, the diffusion of an ion or molecule
with $D \sim 10^{-9}$m$^2$/s over the length of a 1 $\mu$m motor corresponds to an effective speed of 1 mm/s, much larger than the speed of the particle.
To leading order in interaction layer thickness and flux,
$\gamma$ satisfies the Laplace equation with boundary 
condition \hbox{$\Gamma = - {\cal D} \, \hat{n}\cdot\!{\bm \nabla} \gamma$,}
where ${\cal D}$ involves a diffusion coefficient or conductivity.
Then
$\gamma$ is given, up to a constant as ${\cal D}^{-1}{\cal L}\{\Gamma\}$~\cite{Nourhani2015PRE062303}
where ${\cal L}$ is a geometry-dependent Neumann-to-Dirichlet 
operator.
Thus, in the
limit of small P\'eclet number, self-phoretic velocity of a spheroid is
$
\bm{\mathcal{U}}
 = 
-{\mu_{ph} \over 3{\cal VD}} 
\int_{S}
 \hat{n}\cdot {\bm r}
\,{\bm \nabla}_{\!s}{\cal L}\{\Gamma\}
 \, dS 
$.
For axisymmetric flux $\Gamma$, a remarkable 
simplification~\footnote{see Supplementary Information for details of the derivation} 
allows the explicit expression
\begin{align}
& \bm{\mathcal{U}} = 
{\cal U} \, \hat{\bf e}_z =  
\hat{\bf e}_z {(-\mu_{ph}) \over 2 {\cal D}} \int_{-1}^1 \,K(\eta; a/b) \,\Gamma(\eta)  \, d\eta, 
\label{eq:velocity-K-integral}
\end{align}
where, 
as depicted in Fig.~\ref{fig:SpheroidGeo},
$\hat{\bf e}_z$ is the symmetry direction, $a$ ($b$) is the half-length 
along (perpendicular to) the symmetry axis, and   $-1\leq\eta \equiv z/a \leq 1$. 
The dimensionless kernel 
\begin{align}
K(\eta; a/b) = \frac{\eta}{ \sqrt{ \eta^2 + (a/b)^2 (1-\eta^2)} },
 \label{eq:kernelspheroid}
\end{align}
expressing the contribution of flux at each location on the motor surface 
to motion, is the main protagonist of this Letter.

Inspection of the graphs of $K(\eta;a/b)$,
shown in Fig.~\ref{fig:keta} for a range of aspect ratios $a/b$, is already quite revealing.
$K(\eta; \ell^{-1})$ is the reflection of $K(\eta; \ell)$ across the diagonal 
$\ell \equiv a/b =1$. Thus, for a sphere ($a/b =1$), $K$ degenerates to a straight line,
a known result~\cite{Nourhani+-15-PoF}.
Deviation from sphericity by increase of the aspect ratio $a/b$
increasingly suppresses $|K|$ around the equator $\eta=0$.
The earliest generation of self-electrophoretic cylindrical rods~\cite{Paxton:2004p183} had an
aspect ratio of about 5, so this effect is strong even under ordinary conditions.
One message is clear: for a thin rod-like particle,
only surface activity near the poles contributes significantly to self-phoresis.
Thus, in designing a motor with large aspect ratio, details of 
the surface activity around the equatorial region are 
insignificant and may be chosen for convenience. 
This phenomenon explains  why, in numerical simulation of self-electrophoretic 
long rods,  a jump discontinuity in surface cation flux distribution around the equator
can provide the essential physics and give consistent results with experimental 
observations~\cite{Wang:JACS:2012}.

On the contrary, as the particle deviates from sphericity toward a discoidal shape, 
a zone of high effectivity moves from the poles toward the equator. 
This suggests, perhaps rather surprisingly, that oblate designs 
can be much more effective at converting chemical activity into speed.
This is a region of the design space which deserves more experimental attention
than it has received to date. 

Now, we exploit the kernel (\ref{eq:kernelspheroid}) to
explore specific parameterized families of motor designs.
Since simple models without too many parameters are best for revealing generic trends,  
we consider bipartite models with flux taking distinct uniform values over two 
complementary regions of the surface. There are two families to be considered.
Source/inert (or  sink/inert) particles occur 
in cases of self-diffusiophoresis and self-thermophoresis,
with an active region which is a pure source (or sink), and a passive region. 
Self-electrophoretic particles, by contrast, have a source/sink configuration. 
Since the net ion flux from the entire surface must be zero, 
equal quantities of active ions are produced on the source region and consumed on the sink.

The performance characteristic of motor designs in our study
is not simply speed.
Rather, we are interested in the efficacy with which fuel is used to produce speed
as the pattern of flux $\Gamma$ and aspect ratio are varied.
The surface area $S$ serves as a measure of particle size and $\|\Gamma\|$
of total ``activity'' on the particle surface. 
Thus, using the speed scale 
$(-\mu_{ph}/ 2{\cal D})\|\Gamma\| / S$
as a normalization factor (which has no dependence on the spheroidal geometry and
could be used for other shapes)
 to remove sensitivity to uniformly scaling the flux by a constant,
We define the (dimensionless) {\it efficacy}
\begin{align}
\Eff(\Gamma,\ell,S)  
:=
{{\cal U} \over  {(-\mu_{ph}) \over 2{\cal D}}{\|\Gamma\| \over S}}=
{S \over \|\Gamma\|} \int_{-1}^1  K(\eta; \ell) \Gamma(\eta)  \, d\eta.
\label{eq:efficacy}
\end{align}
If we compare motors with the same area $S$ (with $\Gamma$ and $a/b$ variable), 
$\Eff(\Gamma,\ell,S)$ measures, on some scale, the ratio of speed to total activity of the 
motor. This is probably the preferred way to think of it.
In the following we discuss the efficacy of source/since and source/inert configurations.
Closed-form expressions can be found in the Supplementary Information.

\begin{figure}[t]
\begin{center}
\includegraphics[width=0.42\textwidth]{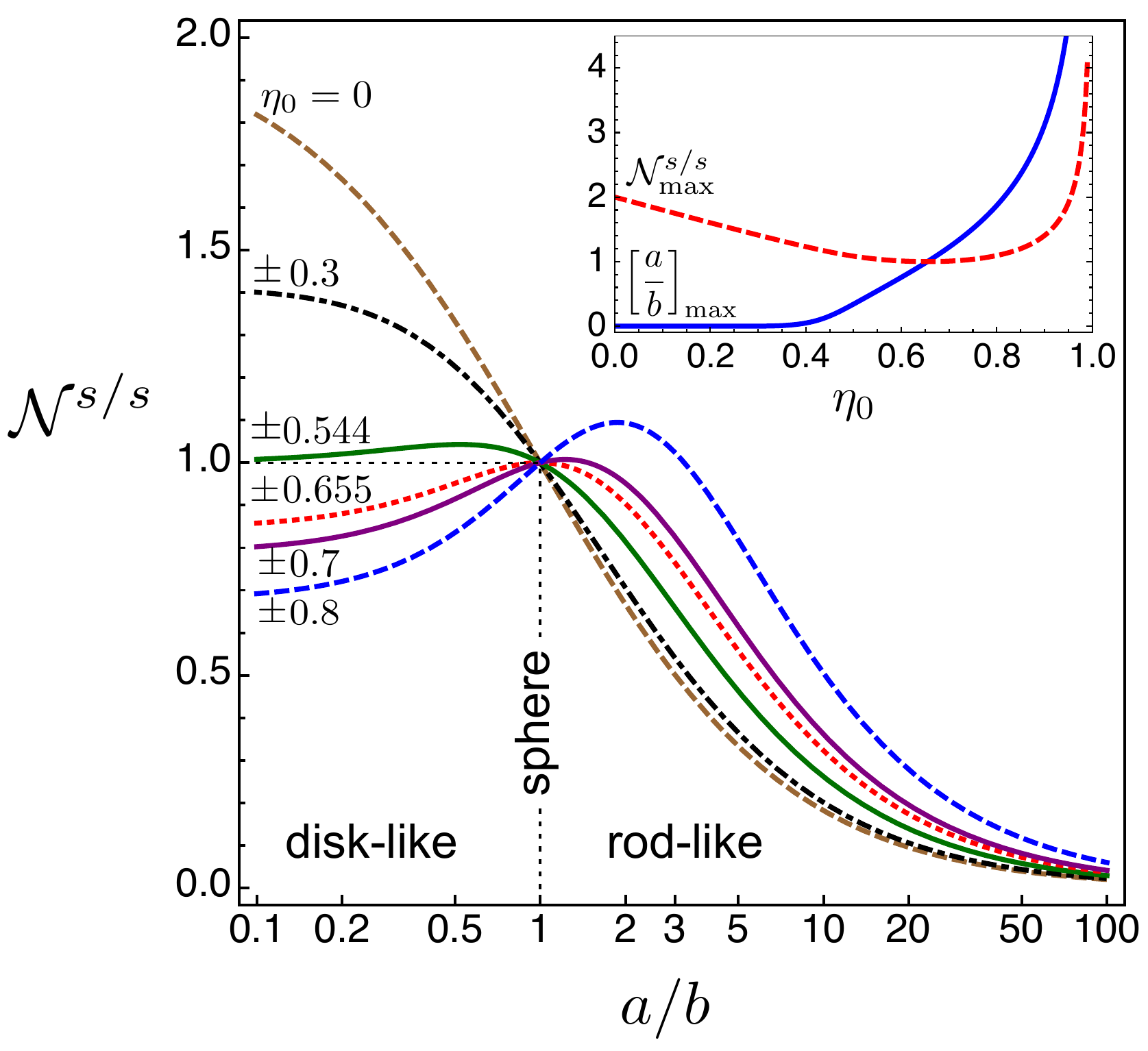}
\end{center}
\vspace{-15pt}
\caption{ 
(color online).
Efficacy of a source/sink motor as a function of aspect ratio for various
values of the source-sink boundary $\eta_0$.
The limit, as aspect ratio tends to zero is
$\lim_{a/b\to 0}\Eff^{s/s} = 2(1-|\eta_0|)/(1-\eta_0^4)$
At $\eta_0=0$, the thin disk limit of the normalized speed is thus 2,
and it exceeds 1 (the spherical value) for $\eta_0 < 0.544$. 
At $\eta_0 = \sqrt{3/7} \approx 0.655$, the derivative at the spherical
point is zero. Below this value, the maximum is for an oblate spheroid, and
above it, for a prolate spheroid, as also shown in the inset.
Inset: 
The dashed red curve shows the maximum attainable value of the 
normalized speed over all aspect ratios for given source/sink boundary
$\eta_0$.
The solid blue curve shows the aspect ratio at which that maximum is
attained. Note, the vertical scale therefore quantifies different
things for the two curves.
The maximizing aspect ratio $[a/b]_{\mathrm{max}}$ ratio is very nearly zero
for all $\eta_0 \lesssim 0.35$.
\label{fig:NGsourcesink}
}
\end{figure}

\noindent{\bf Source/sink.} 
A source/sink design (see Fig. \ref{fig:SpheroidGeo}) 
has a uniform positive flux over the source region
$\{\eta > \eta_0\}$ of area $S_+$ and a uniform negative flux over
the sink region $\{\eta > \eta_0\}$ of area $S_- = S - S_+$.
From the definition (\ref{eq:efficacy}), and the constraint of
zero net flux from the surface, the efficacy for this case is (superscript ``$s/s$'' indicates ``source/sink'')
 \begin{equation}
\label{eq:speed-source-sink}
{\Eff}^{s/s}(\eta_0,\ell) = \frac{ S^2 }{2 S_- S_+}  
\int_{|\eta_0|}^1 K(\eta;\ell)d\eta,
 \end{equation}
 which is even in $\eta_0$.
Figure~\ref{fig:NGsourcesink} shows $\Eff^{s/s}(\eta_0,a/b)$ for a variety
of $\eta_0$ values as a function of aspect ratio.
In the discoidal regime, speed {\em decreases} monotonically
as $|\eta_0|$ is increased, whereas in the prolate regime, it {\em increases}
monotonically. 
Note that this involves comparing particles of identical size
and shape, so that the same trend is valid if total flux
is held fixed.
For $\eta_0 = 0$, the source/sink efficacy
$\Eff^{s/s}$ tends to 2 as $a/b \to 0$.
For small values of $\eta_0$ oblate designs have highest efficacy.
The relative advantage of discoidal designs decreases
as $\eta_0$ increases. At $\eta_0 \approx 0.655$, the maximum 
efficacy is for a sphere, and beyond that, it lies in the prolate range.
For $\eta_0 \gtrsim 0.655$, then, for every oblate shape ($a/b < 1$), 
there is an equally  efficacious prolate shape.
The efficacy of a sphere is completely insensitive to $\eta_0$,
a fact noted previously~\cite{Nourhani+-15-PoF}. 
Now we see that this feature is specific to the sphere.
The aspect ratio $[a/b]_{\text{max}}$ which maximizes 
$\Eff^{s/s}(\eta_0,a/b)$ at fixed $\eta_0$ 
and the corresponding maximum value $\Eff^{s/s}(\eta_0,[a/b]_{\text{max}})$
are plotted in the inset to Fig.~\ref{fig:NGsourcesink}
as the red dashed and blue solid curves, respectively.
Only non-negative
$\eta_0$ are shown since $\Eff^{s/s}(-\eta_0,\ell) = \Eff^{s/s}(\eta_0,\ell)$.
Over most of the range of $\eta_0$, the variation is no more
than a factor of two. As $\eta_0 \to 1$, $\Eff^{s/s}_{\mathrm{max}}$
diverges, but this limit is probably not realistic. The required
shape becomes too
elongated, and the large imbalance
in source and sink areas implies that kinetics will strongly limit
the attainable flux.
\begin{figure}[t]
  \begin{center}
    \includegraphics[width=0.36\textwidth]{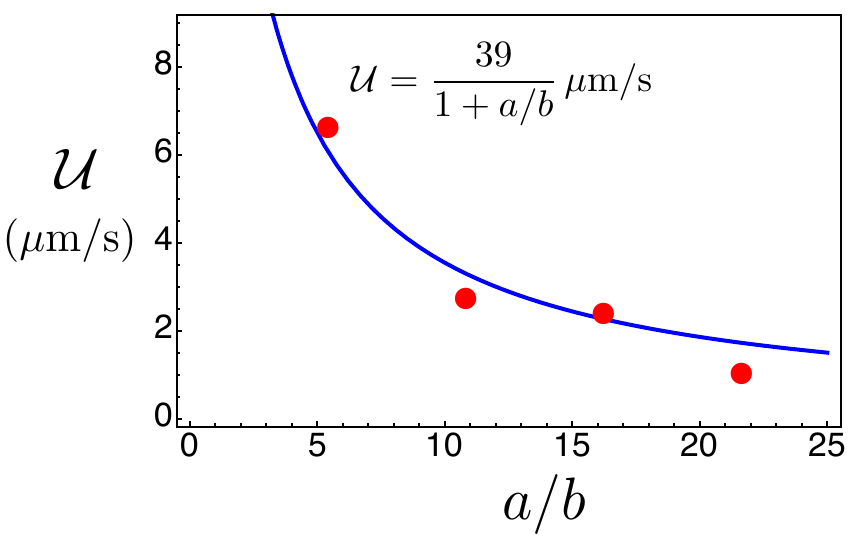}
  \end{center}
\caption{
\label{fig:Dhar-fit}
(color online). Fit of speed measurements\cite{Dhar:NL:2006} on half Au, half Pt cylinders of 
varying length to the formula ${\cal U}= 2(-\mu_{ph}\alpha/2{\cal D})/(1+a/b)$ 
derived for $\eta_0 = 0$ spheroids.
The single fitting parameter is $-\mu_{ph}\alpha/2{\cal D} \simeq 19.5$ $\mu$m/s,
which is the velocity scale characterizing these motors. 
}
\end{figure}

For the common antisymmetric design, $\eta_0~=~0$, Eq.~(\ref{eq:speed-source-sink}) 
yields the simple explicit expression 
\hbox{${\cal U} = 2(-\mu_{ph}\alpha/2{\cal D})/(1+a/b)$}, with 
$\alpha$ the uniform value of $\Gamma$ on the source.
Dhar {\it et al.}~\cite{Dhar:NL:2006} measured speeds of cylindrical 
half-Au/half-Pt rods of the same radius but differing diameters in the 
same medium. 
Approximating those shapes by spheroids and applying the source/sink model
with the same value of $\alpha$ for all rods leaves a single fitting parameter.
{The fit in Fig.~\ref{fig:Dhar-fit} seems good enough that if data were available 
for a variety of fuel concentrations, it might be possible to discover useful 
information about the reaction kinetics.

The simple formula in the previous paragraph is {\em exact} for all
spheroids. 
Some expressions for the slender body limit $a/b\gg 1$
which have appeared in the literature~\cite{Golestanian+-07,Popescu2010EPJ,Paxton:2004p183}
contain logarithmic factors, which can now be seen to be spurious, the correct
asymptotic behavior being simply $b/a$. This has also been noted 
in ~\cite{SchnitzerYariv2105}.

\noindent{\bf Source/inert or sink/inert.}
The difference between a pure source and a pure sink configuration is a
simple matter of sign, so we consider just the source case, with 
uniform source over the region $\{\eta>\eta_0\}$ of area $S_+$; 
the rest of the motor surface is inert (Fig.~\ref{fig:SpheroidGeo}).
Equation~(\ref{eq:efficacy}) yields 
(``$s/i$'' superscript indicates ``source/inert'')
  \begin{equation}
\label{eq:speed-source-inert}
\Eff^{s/i}(\eta_0,\ell) =  \frac{S}{S_+}  
\int_{|\eta_0|}^1 K(\eta;\ell)d\eta.
 \end{equation}
We concentrate now on differences from the source/sink geometry.
In contrast to that case, ${\cal N}^{s/i}$ is not even in $\eta_0$,
since there is no symmetry to guarantee that.
Consequently, Fig.~\ref{fig:source} shows a range for both
small source ($\eta_0 > 0$) and  large source ($\eta_0 < 0$).
For $\eta_0 > 0$, the curves are qualitatively similar to those for a source/sink
particle. 
For $\eta_0 < 0$, however, the two cases differ greatly. 
In particular, the efficacy is a monotonically decreasing function of aspect 
ratio for all $\eta_0 \leq 0$.
In the spherical case (${a/b = 1}$), the efficacy
$\Eff^{s/i}(\eta_0,0) = 1+\eta_0$ is not independent of $\eta_0$ as it
was for a source/sink particle, but decreases with  increase in
the source area.
In the extreme discotic limit $a\ll b$, the motor efficacy 
$\Eff^{s/i} = 2\,(1-|\eta_0|)/(1-\eta_0|\eta_0|)$ has a maximum value
$2$ at the half-and-half geometry $\eta_0 = 0$, as for a source/sink
particle, and decreases as $|\eta_0|$ increases.
The maximum attainable efficacy at fixed $\eta_0$ and the
corresponding value of aspect ratio at which it is attained are
also qualitatively similar to the source/sink case for $\eta_0 > 0$.
However,  as $\eta_0$ decreases from zero to $-1$ 
the maximum is
attained only in the limit $a/b \to 0$
and $\Eff^{s/i}_\text{max}$ drops monotonically from 2 to zero,
because, as $\eta_0$ becomes
more and more negative, the source is spread more and more evenly across 
the surface so that activity at one end counteracts that at the other.
The speed (and not the efficacy) for constant surface activity $\|\Gamma\|/S_+$ is always maximum for 
any value of $\eta_0$ for extreme oblate 
particles~\footnote{see Sect. 3 of Supplementary Information}
and it is previously observed for $\eta_0=0$~\cite{Popescu2010EPJ}.
%

\begin{figure}[t]
  \begin{center}
    \includegraphics[width=0.42\textwidth]{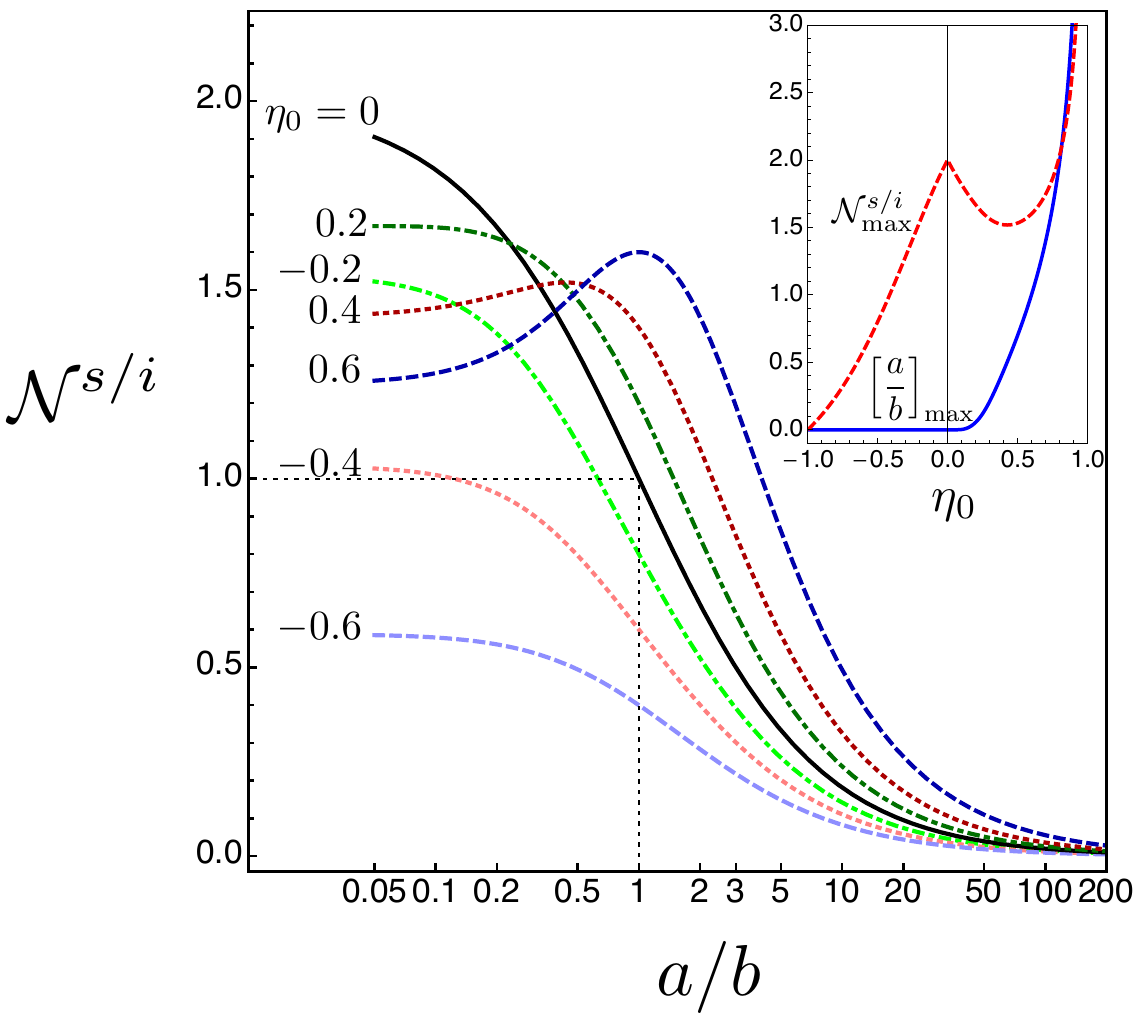}
  \end{center}
\caption{
(color online).
Efficacy of a pure-source motor as a function of aspect ratio for various
values of the source-inert boundary $\eta_0$.
Inset: Maximum attainable efficacy at fixed $\eta_0$ (dashed red),
and the aspect ratio at which it is attained (solid blue).
Compare Fig. \ref{fig:NGsourcesink} for the source/sink case.}
\label{fig:source}
\end{figure}

Understanding the effect of geometry and surface activity on autonomous active colloids
is essential for designing powered machines with tuned properties at nano- and micro-scale. 
While in phoresis of passive particles the driving field is external, active colloids harvest energy from their 
environment and self-generate the driving filed through a surface flux. 
The self-phoretic velocity expression (\ref{eq:velocity-K-integral}),
connecting the motor velocity to its shape and distribution of surface flux,
is general across the various self-phoretic mechanisms, opening the route to
a unified formulation connecting velocity to surface activity and flux for various geometries
 with uniform phoretic mobility.
For an arbitrary axisymmetric geometry with  axisymmetric flux,
in consequence of the linearity and  scaling properties of the governing 
equations, a formula like (\ref{eq:velocity-K-integral}) holds with some kernel $K$
expressing the contribution of flux at each location on the motor surface to 
motion.
The spheroid family distinguishes itself through the
explicit flux-to-speed kernel (\ref{eq:kernelspheroid}).
Since phoretic mobility $\mu_{ph}$ and diffusivity ${\cal D}$ are 
constant material properties, dimensional analysis then shows that 
$K$ is dimensionless. Therefore, so long as $\Gamma(\eta)$ is held 
constant,  the size of the particle drops out.
For a given self-phoresis mechanism, $\Gamma(\eta)$ may have an implicit size dependence when directly 
controllable conditions such as fuel concentration are maintained constant~\cite{Nourhani2015PRE062303}.
However, such dependences vary from one self-phoretic mechanism to another, 
and thus fall outside our unifying scope. For unifying studies across different 
self-phoretic mechanisms, we regard $\Gamma(\eta)$ as simply given in 
expression (\ref{eq:velocity-K-integral}).

Even with simple assumptions about the surface flux distribution,
the theory make good contact with experimental results,
as Fig. \ref{fig:Dhar-fit} shows,
and explains the consistency of experimental observation with numerical 
simulation result for a discontinuous flux jump around the equator in Ref.~\cite{Wang:JACS:2012}.
Oblate (discotic) spheroids --- to our knowledge these have received no previous 
experimental attention --- emerge from this survey of the design space as 
potentially interesting candidates for experimental investigation.

\begin{acknowledgments}
The authors are grateful to Prof. Vincent H. Crespi, Prof. Wei Wang and Dr. Cristiano Nisoli for 
their insightful comments and suggestions.
This work was funded by the Penn State MRSEC, Center for Nanoscale Science, under 
award National Science Foundation DMR-1420620.
\end{acknowledgments}

\begin{thebibliography}{22}%
\makeatletter
\providecommand \@ifxundefined [1]{%
 \@ifx{#1\undefined}
}%
\providecommand \@ifnum [1]{%
 \ifnum #1\expandafter \@firstoftwo
 \else \expandafter \@secondoftwo
 \fi
}%
\providecommand \@ifx [1]{%
 \ifx #1\expandafter \@firstoftwo
 \else \expandafter \@secondoftwo
 \fi
}%
\providecommand \natexlab [1]{#1}%
\providecommand \enquote  [1]{``#1''}%
\providecommand \bibnamefont  [1]{#1}%
\providecommand \bibfnamefont [1]{#1}%
\providecommand \citenamefont [1]{#1}%
\providecommand \href@noop [0]{\@secondoftwo}%
\providecommand \href [0]{\begingroup \@sanitize@url \@href}%
\providecommand \@href[1]{\@@startlink{#1}\@@href}%
\providecommand \@@href[1]{\endgroup#1\@@endlink}%
\providecommand \@sanitize@url [0]{\catcode `\\12\catcode `\$12\catcode
  `\&12\catcode `\#12\catcode `\^12\catcode `\_12\catcode `\%12\relax}%
\providecommand \@@startlink[1]{}%
\providecommand \@@endlink[0]{}%
\providecommand \url  [0]{\begingroup\@sanitize@url \@url }%
\providecommand \@url [1]{\endgroup\@href {#1}{\urlprefix }}%
\providecommand \urlprefix  [0]{URL }%
\providecommand \Eprint [0]{\href }%
\providecommand \doibase [0]{http://dx.doi.org/}%
\providecommand \selectlanguage [0]{\@gobble}%
\providecommand \bibinfo  [0]{\@secondoftwo}%
\providecommand \bibfield  [0]{\@secondoftwo}%
\providecommand \translation [1]{[#1]}%
\providecommand \BibitemOpen [0]{}%
\providecommand \bibitemStop [0]{}%
\providecommand \bibitemNoStop [0]{.\EOS\space}%
\providecommand \EOS [0]{\spacefactor3000\relax}%
\providecommand \BibitemShut  [1]{\csname bibitem#1\endcsname}%
\let\auto@bib@innerbib\@empty
\bibitem [{\citenamefont {Ebbens}\ and\ \citenamefont
  {Howse}(2010)}]{Ebbens:2010p86}%
  \BibitemOpen
  \bibfield  {author} {\bibinfo {author} {\bibfnamefont {S.~J.}\ \bibnamefont
  {Ebbens}}\ and\ \bibinfo {author} {\bibfnamefont {J.~R.}\ \bibnamefont
  {Howse}},\ }\href {\doibase 10.1039/b918598d} {\bibfield  {journal} {\bibinfo
   {journal} {Soft Matter}\ }\textbf {\bibinfo {volume} {6}},\ \bibinfo {pages}
  {726} (\bibinfo {year} {2010})}\BibitemShut {NoStop}%
\bibitem [{\citenamefont {Wang}(2013)}]{JWangBook2013}%
  \BibitemOpen
  \bibfield  {author} {\bibinfo {author} {\bibfnamefont {J.}~\bibnamefont
  {Wang}},\ }\href@noop {} {\emph {\bibinfo {title} {Nanomachines: Fundamentals
  and Applications}}}\ (\bibinfo  {publisher} {Wiley-VCH},\ \bibinfo {year}
  {2013})\BibitemShut {NoStop}%
\bibitem [{\citenamefont {Paxton}\ \emph {et~al.}(2004)\citenamefont {Paxton},
  \citenamefont {Kistler}, \citenamefont {Olmeda}, \citenamefont {Sen},
  \citenamefont {Angelo}, \citenamefont {Cao}, \citenamefont {Mallouk},
  \citenamefont {Lammert},\ and\ \citenamefont {Crespi}}]{Paxton:2004p183}%
  \BibitemOpen
  \bibfield  {author} {\bibinfo {author} {\bibfnamefont {W.}~\bibnamefont
  {Paxton}}, \bibinfo {author} {\bibfnamefont {K.}~\bibnamefont {Kistler}},
  \bibinfo {author} {\bibfnamefont {C.}~\bibnamefont {Olmeda}}, \bibinfo
  {author} {\bibfnamefont {A.}~\bibnamefont {Sen}}, \bibinfo {author}
  {\bibfnamefont {S.~S.}\ \bibnamefont {Angelo}}, \bibinfo {author}
  {\bibfnamefont {Y.}~\bibnamefont {Cao}}, \bibinfo {author} {\bibfnamefont
  {T.}~\bibnamefont {Mallouk}}, \bibinfo {author} {\bibfnamefont
  {P.}~\bibnamefont {Lammert}}, \ and\ \bibinfo {author} {\bibfnamefont
  {V.}~\bibnamefont {Crespi}},\ }\href {\doibase 10.1021/ja047697z} {\bibfield
  {journal} {\bibinfo  {journal} {J Am Chem Soc}\ }\textbf {\bibinfo {volume}
  {126}},\ \bibinfo {pages} {13424} (\bibinfo {year} {2004})}\BibitemShut
  {NoStop}%
\bibitem [{\citenamefont {Gibbs}\ and\ \citenamefont
  {Fischer}(2015)}]{C5CC00565E}%
  \BibitemOpen
  \bibfield  {author} {\bibinfo {author} {\bibfnamefont {J.~G.}\ \bibnamefont
  {Gibbs}}\ and\ \bibinfo {author} {\bibfnamefont {P.}~\bibnamefont
  {Fischer}},\ }\href {\doibase 10.1039/C5CC00565E} {\bibfield  {journal}
  {\bibinfo  {journal} {Chem. Commun.}\ }\textbf {\bibinfo {volume} {51}},\
  \bibinfo {pages} {4192} (\bibinfo {year} {2015})}\BibitemShut {NoStop}%
\bibitem [{\citenamefont {Popescu}\ \emph {et~al.}(2010)\citenamefont
  {Popescu}, \citenamefont {Dietrich}, \citenamefont {Tasinkevych},\ and\
  \citenamefont {Ralston}}]{Popescu2010EPJ}%
  \BibitemOpen
  \bibfield  {author} {\bibinfo {author} {\bibfnamefont {M.~N.}\ \bibnamefont
  {Popescu}}, \bibinfo {author} {\bibfnamefont {S.}~\bibnamefont {Dietrich}},
  \bibinfo {author} {\bibfnamefont {M.}~\bibnamefont {Tasinkevych}}, \ and\
  \bibinfo {author} {\bibfnamefont {J.}~\bibnamefont {Ralston}},\ }\href@noop
  {} {\bibfield  {journal} {\bibinfo  {journal} {Eur. Phys. J. E}\ }\textbf
  {\bibinfo {volume} {31}},\ \bibinfo {pages} {351} (\bibinfo {year}
  {2010})}\BibitemShut {NoStop}%
\bibitem [{\citenamefont {Golestanian}\ \emph {et~al.}(2007)\citenamefont
  {Golestanian}, \citenamefont {Liverpool},\ and\ \citenamefont
  {Ajdari}}]{Golestanian+-07}%
  \BibitemOpen
  \bibfield  {author} {\bibinfo {author} {\bibfnamefont {R.}~\bibnamefont
  {Golestanian}}, \bibinfo {author} {\bibfnamefont {T.~B.}\ \bibnamefont
  {Liverpool}}, \ and\ \bibinfo {author} {\bibfnamefont {A.}~\bibnamefont
  {Ajdari}},\ }\href@noop {} {\bibfield  {journal} {\bibinfo  {journal} {New
  Journal of Physics}\ }\textbf {\bibinfo {volume} {9}} (\bibinfo {year}
  {2007})}\BibitemShut {NoStop}%
\bibitem [{\citenamefont {Nourhani}\ \emph
  {et~al.}(2015{\natexlab{a}})\citenamefont {Nourhani}, \citenamefont
  {Crespi},\ and\ \citenamefont {Lammert}}]{Nourhani2015PRE062303}%
  \BibitemOpen
  \bibfield  {author} {\bibinfo {author} {\bibfnamefont {A.}~\bibnamefont
  {Nourhani}}, \bibinfo {author} {\bibfnamefont {V.~H.}\ \bibnamefont
  {Crespi}}, \ and\ \bibinfo {author} {\bibfnamefont {P.~E.}\ \bibnamefont
  {Lammert}},\ }\href@noop {} {\bibfield  {journal} {\bibinfo  {journal} {Phys.
  Rev. E}\ }\textbf {\bibinfo {volume} {91}},\ \bibinfo {pages} {062303}
  (\bibinfo {year} {2015}{\natexlab{a}})}\BibitemShut {NoStop}%
\bibitem [{\citenamefont {Nourhani}\ \emph
  {et~al.}(2015{\natexlab{b}})\citenamefont {Nourhani}, \citenamefont {Crespi},
  \citenamefont {Lammert},\ and\ \citenamefont
  {Borhan}}]{Nourhani+-15-PoF-spheroid}%
  \BibitemOpen
  \bibfield  {author} {\bibinfo {author} {\bibfnamefont {A.}~\bibnamefont
  {Nourhani}}, \bibinfo {author} {\bibfnamefont {V.~H.}\ \bibnamefont
  {Crespi}}, \bibinfo {author} {\bibfnamefont {P.~E.}\ \bibnamefont {Lammert}},
  \ and\ \bibinfo {author} {\bibfnamefont {A.}~\bibnamefont {Borhan}},\
  }\href@noop {} {\bibfield  {journal} {\bibinfo  {journal} {Phys Fluids}\
  }\textbf {\bibinfo {volume} {27}},\ \bibinfo {pages} {092002} (\bibinfo
  {year} {2015}{\natexlab{b}})}\BibitemShut {NoStop}%
\bibitem [{\citenamefont {Nourhani}\ \emph
  {et~al.}(2015{\natexlab{c}})\citenamefont {Nourhani}, \citenamefont
  {Lammert}, \citenamefont {Crespi},\ and\ \citenamefont
  {Borhan}}]{Nourhani+-15-PoF}%
  \BibitemOpen
  \bibfield  {author} {\bibinfo {author} {\bibfnamefont {A.}~\bibnamefont
  {Nourhani}}, \bibinfo {author} {\bibfnamefont {P.~E.}\ \bibnamefont
  {Lammert}}, \bibinfo {author} {\bibfnamefont {V.~H.}\ \bibnamefont {Crespi}},
  \ and\ \bibinfo {author} {\bibfnamefont {A.}~\bibnamefont {Borhan}},\
  }\href@noop {} {\bibfield  {journal} {\bibinfo  {journal} {Phys Fluids}\
  }\textbf {\bibinfo {volume} {27}},\ \bibinfo {pages} {012001} (\bibinfo
  {year} {2015}{\natexlab{c}})}\BibitemShut {NoStop}%
\bibitem [{\citenamefont {Ebbens}\ \emph {et~al.}(2012)\citenamefont {Ebbens},
  \citenamefont {Tu}, \citenamefont {Howse},\ and\ \citenamefont
  {Golestanian}}]{EbbensPRERp020401}%
  \BibitemOpen
  \bibfield  {author} {\bibinfo {author} {\bibfnamefont {S.}~\bibnamefont
  {Ebbens}}, \bibinfo {author} {\bibfnamefont {M.-H.}\ \bibnamefont {Tu}},
  \bibinfo {author} {\bibfnamefont {J.~R.}\ \bibnamefont {Howse}}, \ and\
  \bibinfo {author} {\bibfnamefont {R.}~\bibnamefont {Golestanian}},\
  }\href@noop {} {\bibfield  {journal} {\bibinfo  {journal} {Phys. Rev. E (R)}\
  }\textbf {\bibinfo {volume} {85}},\ \bibinfo {pages} {020401} (\bibinfo
  {year} {2012})}\BibitemShut {NoStop}%
\bibitem [{\citenamefont {Schnitzer}\ and\ \citenamefont
  {Yariv}(2015)}]{SchnitzerYariv2105}%
  \BibitemOpen
  \bibfield  {author} {\bibinfo {author} {\bibfnamefont {O.}~\bibnamefont
  {Schnitzer}}\ and\ \bibinfo {author} {\bibfnamefont {E.}~\bibnamefont
  {Yariv}},\ }\href@noop {} {\bibfield  {journal} {\bibinfo  {journal} {Phys
  Fluids}\ }\textbf {\bibinfo {volume} {27}},\ \bibinfo {pages} {031701}
  (\bibinfo {year} {2015})}\BibitemShut {NoStop}%
\bibitem [{\citenamefont {Yariv}(2011)}]{Yariv:2011p74}%
  \BibitemOpen
  \bibfield  {author} {\bibinfo {author} {\bibfnamefont {E.}~\bibnamefont
  {Yariv}},\ }\href {\doibase 10.1098/rspa.2010.0503} {\bibfield  {journal}
  {\bibinfo  {journal} {Proceedings of the Royal Society A: Mathematical,
  Physical and Engineering Sciences}\ }\textbf {\bibinfo {volume} {467}},\
  \bibinfo {pages} {1645} (\bibinfo {year} {2011})}\BibitemShut {NoStop}%
\bibitem [{\citenamefont {Sabass}\ and\ \citenamefont
  {Seifert}(2012)}]{Sabass:JCP:2012}%
  \BibitemOpen
  \bibfield  {author} {\bibinfo {author} {\bibfnamefont {B.}~\bibnamefont
  {Sabass}}\ and\ \bibinfo {author} {\bibfnamefont {U.}~\bibnamefont
  {Seifert}},\ }\href@noop {} {\bibfield  {journal} {\bibinfo  {journal} {J.
  Chem. Phys.}\ }\textbf {\bibinfo {volume} {136}},\ \bibinfo {pages} {214507}
  (\bibinfo {year} {2012})}\BibitemShut {NoStop}%
\bibitem [{\citenamefont {Jiang}\ \emph {et~al.}(2010)\citenamefont {Jiang},
  \citenamefont {Yoshinaga},\ and\ \citenamefont {Sano}}]{Jiang2010PRL268302}%
  \BibitemOpen
  \bibfield  {author} {\bibinfo {author} {\bibfnamefont {H.~R.}\ \bibnamefont
  {Jiang}}, \bibinfo {author} {\bibfnamefont {N.}~\bibnamefont {Yoshinaga}}, \
  and\ \bibinfo {author} {\bibfnamefont {M.}~\bibnamefont {Sano}},\ }\href@noop
  {} {\bibfield  {journal} {\bibinfo  {journal} {Phys Rev. Lett.}\ }\textbf
  {\bibinfo {volume} {105}},\ \bibinfo {pages} {268302} (\bibinfo {year}
  {2010})}\BibitemShut {NoStop}%
\bibitem [{\citenamefont {Golestanian}(2012)}]{Golestanian2012PRL038303}%
  \BibitemOpen
  \bibfield  {author} {\bibinfo {author} {\bibfnamefont {R.}~\bibnamefont
  {Golestanian}},\ }\href@noop {} {\bibfield  {journal} {\bibinfo  {journal}
  {Phys Rev. Lett.}\ }\textbf {\bibinfo {volume} {108}},\ \bibinfo {pages}
  {038303} (\bibinfo {year} {2012})}\BibitemShut {NoStop}%
\bibitem [{\citenamefont {Reigh}\ and\ \citenamefont
  {Kapral}(2015)}]{Reigh+Kapral-15}%
  \BibitemOpen
  \bibfield  {author} {\bibinfo {author} {\bibfnamefont {S.~Y.}\ \bibnamefont
  {Reigh}}\ and\ \bibinfo {author} {\bibfnamefont {R.}~\bibnamefont {Kapral}},\
  }\href {\doibase 10.1039/c4sm02857k} {\bibfield  {journal} {\bibinfo
  {journal} {Soft Matter}\ }\textbf {\bibinfo {volume} {11}},\ \bibinfo {pages}
  {3149} (\bibinfo {year} {2015})}\BibitemShut {NoStop}%
\bibitem [{\citenamefont {Anderson}(1989)}]{Anderson-89}%
  \BibitemOpen
  \bibfield  {author} {\bibinfo {author} {\bibfnamefont {J.~L.}\ \bibnamefont
  {Anderson}},\ }\href@noop {} {\bibfield  {journal} {\bibinfo  {journal}
  {Annual Review of Fluid Mechanics}\ }\textbf {\bibinfo {volume} {21}},\
  \bibinfo {pages} {61} (\bibinfo {year} {1989})}\BibitemShut {NoStop}%
\bibitem [{\citenamefont {Fair}\ and\ \citenamefont
  {Anderson}(1989)}]{Fair+Anderson}%
  \BibitemOpen
  \bibfield  {author} {\bibinfo {author} {\bibfnamefont {M.~C.}\ \bibnamefont
  {Fair}}\ and\ \bibinfo {author} {\bibfnamefont {J.~L.}\ \bibnamefont
  {Anderson}},\ }\href {\doibase 10.1016/0021-9797(89)90045-3} {\bibfield
  {journal} {\bibinfo  {journal} {Journal Of Colloid And Interface Science}\
  }\textbf {\bibinfo {volume} {127}},\ \bibinfo {pages} {388} (\bibinfo {year}
  {1989})}\BibitemShut {NoStop}%
\bibitem [{Note1()}]{Note1}%
  \BibitemOpen
  \bibinfo {note} {See Supplementary Information for details of the
  derivation}\BibitemShut {NoStop}%
\bibitem [{\citenamefont {Wang}\ \emph {et~al.}(2013)\citenamefont {Wang},
  \citenamefont {Chiang}, \citenamefont {Velegol},\ and\ \citenamefont
  {Mallouk}}]{Wang:JACS:2012}%
  \BibitemOpen
  \bibfield  {author} {\bibinfo {author} {\bibfnamefont {W.}~\bibnamefont
  {Wang}}, \bibinfo {author} {\bibfnamefont {T.-Y.}\ \bibnamefont {Chiang}},
  \bibinfo {author} {\bibfnamefont {D.}~\bibnamefont {Velegol}}, \ and\
  \bibinfo {author} {\bibfnamefont {T.~E.}\ \bibnamefont {Mallouk}},\
  }\href@noop {} {\bibfield  {journal} {\bibinfo  {journal} {J Am Chem Soc}\
  }\textbf {\bibinfo {volume} {135}},\ \bibinfo {pages} {10557} (\bibinfo
  {year} {2013})}\BibitemShut {NoStop}%
\bibitem [{\citenamefont {Dhar}\ \emph {et~al.}(2006)\citenamefont {Dhar},
  \citenamefont {Fischer}, \citenamefont {Wang}, \citenamefont {Mallouk},
  \citenamefont {Paxton},\ and\ \citenamefont {Sen}}]{Dhar:NL:2006}%
  \BibitemOpen
  \bibfield  {author} {\bibinfo {author} {\bibfnamefont {P.}~\bibnamefont
  {Dhar}}, \bibinfo {author} {\bibfnamefont {T.~M.}\ \bibnamefont {Fischer}},
  \bibinfo {author} {\bibfnamefont {Y.}~\bibnamefont {Wang}}, \bibinfo {author}
  {\bibfnamefont {T.~E.}\ \bibnamefont {Mallouk}}, \bibinfo {author}
  {\bibfnamefont {W.~F.}\ \bibnamefont {Paxton}}, \ and\ \bibinfo {author}
  {\bibfnamefont {A.}~\bibnamefont {Sen}},\ }\href@noop {} {\bibfield
  {journal} {\bibinfo  {journal} {Nano Lett}\ }\textbf {\bibinfo {volume}
  {6}},\ \bibinfo {pages} {66} (\bibinfo {year} {2006})}\BibitemShut {NoStop}%
\bibitem [{Note2()}]{Note2}%
  \BibitemOpen
  \bibinfo {note} {See Sect. 3 of Supplementary Information}\BibitemShut
  {NoStop}%
\end{thebibliography}

%

\end{document}


\renewcommand{\bottomfraction}{0.8}
\renewcommand{\topfraction}{0.8}
\renewcommand{\textfraction}{0.2}
\renewcommand{\floatpagefraction}{0.8}
\renewcommand{\thesection}{\arabic{section}}



\newcommand{\beq}{\begin{equation}}
\newcommand{\eeq}{\end{equation}}
\newcommand{\eqan}{\begin{eqnarray*}}
\newcommand{\enan}{\end{eqnarray*}}
\newcommand{\spz}{\hspace{0.7cm}}
\newcommand{\lbl}{\label}

\newcommand{\bjerrum}{\lambda_{_{\rm B}}}

\def\Bbb{\mathbb}


\newcommand{\Dslash}{{\slash{\kern -0.5em}\partial}}
\newcommand{\Aslash}{{\slash{\kern -0.5em}A}}

\def\sqr#1#2{{\vcenter{\hrule height.#2pt
     \hbox{\vrule width.#2pt height#1pt \kern#1pt
        \vrule width.#2pt}
     \hrule height.#2pt}}}
\def\smallsquare{\mathchoice\sqr34\sqr34\sqr{2.1}3\sqr{1.5}3}
\def\square{\mathchoice\sqr68\sqr68\sqr{4.2}6\sqr{3.0}6}
 
\def\thinspace{\kern .16667em}
\def\punto{\thinspace .\thinspace}
 
\def\xp{x_{{\kern -.2em}_\perp}}
\def\subp{_{{\kern -.2em}_\perp}}
\def\kperp{k\subp}

\def\derpp#1#2{{\partial #1\over\partial #2}}
\def\derp#1{{\partial~\over\partial #1}}

\def\kap{\overline{\kappa}}
\def\nnabla{\tilde{\nabla}}

\renewcommand{\thetable}{S\arabic{table}}   
\renewcommand{\thefigure}{S\arabic{figure}}
\renewcommand{\theequation}{S\arabic{equation}}

\title{Supplementary Information for \\ ``Efficacy of self-phoretic colloids and microswimmers''}

\author{Amir Nourhani}
\author{Paul E. Lammert}
\affiliation{Center for Nanoscale Science,The Pennsylvania State University, University Park, PA 16802}
\affiliation{Department of Physics, The Pennsylvania State University, University Park, PA 16802}


\maketitle

\newcommand{\ecc}{\varepsilon}
\newcommand{\sph}{{\mathsf s}}
\newcommand{\NtD}{{\mathcal L}}
\newcommand{\thin}{{\frak b}}
\newcommand{\arc}{\ell}


Section \ref{sec:derivation} provides the derivation of the fundamental integral
kernel $K(\eta;\ell)$. Section \ref{sec:closed-form} provides the closed form
expressions needed to evaluate velocities and efficacies. Section \ref{sec:Popescu}
gives supplementary results for the velocity of the diffusiophoretic model of
Popescu {\it et al.}~[5]

\section{Derivation of Kernel}
\label{sec:derivation}

Here, we obtain the integration kernel $K(\eta, a/b)$ for Eq.~(1) in the manuscript,
\begin{align}
& \bm{\mathcal{U}} = \hat{\bf e}_z {(- \mu_{ph}) \over 2 {\cal D}} \int_{-1}^1 \,K(\eta, a/b) \,\Gamma(\eta)  \, d\eta. 
\label{eq:velocity-K-integral}
\end{align}
Adjacent to the motor suface the fluid is subjected to a slip velocity 
${\bm v}_\text{slip} = \mu_{ph} {\bm \nabla}_{\!s}\gamma$ 
where $\gamma$ is harmonic ($\nabla^2\gamma = 0$) with far-field 
constant value $\gamma_\infty$.  The particle is a spheroid of semi-major
axis $a$ and semi-minor axis $b$ with volume ${\cal V} = {4\over3}\pi b^2a$.
According to the self-consistent nonlocal feedback (SCNLF) theory~[7], 
the field $\gamma$ 
can be obtained from the axisymmetric flux $\Gamma$ using 
a geometry-dependent Neumann-to-Dirichlet operator ${\cal L}$.
That is, 
\beq
\Gamma = - {\cal D} \, \hat{n}\cdot\!{\bm \nabla} \gamma
\qquad \leadsto \qquad
({\cal D}^{-1}\Gamma) = -  \, \hat{n}\cdot\!{\bm \nabla} \gamma
\qquad \leadsto \qquad
\gamma = \gamma_\infty + {\cal L}\{{\cal D}^{-1}\Gamma\}. 
\eeq
Since the geometry and flux $\Gamma$ are axisymmetric, the self-phoretic particle 
moves along its symmetry axis $\hat{\bf e}_z$. 
We employ the prolate spheroidal coordinates with the metric factors 
 \begin{align}
g_1(\xi,\eta) =  \xi_s \sqrt{(\xi^2 - 1)/ (\xi^2 - \eta^2)},
  \qquad 
 g_2(\xi,\eta) = \xi_s \sqrt{(1 - \eta^2)/ (\xi^2 - \eta^2)},
\end{align}
which, on the spheroid's surface $\xi_s = \varepsilon^{-1}$, turn into
 \begin{align}
  \varepsilon g_{1,s}(\eta) := \varepsilon g_1(\xi_s,\eta) = {1\over a} \hat{n}\cdot{\bm r}= \sqrt{1 - \varepsilon^2 \over 1 - \varepsilon^2 \eta^2},
  \qquad 
g_2(\eta) := g_{2,s}(\xi_s,\eta) = \hat{\bf e}_\eta \cdot \hat{\bf e}_z= \sqrt{1 - \eta^2 \over1 - \varepsilon^2 \eta^2}.
\end{align}

We also need the surface gradient operator
${\nabla}_{\!s} = \hat{\bf e}_\eta\, a^{-1} g_{2,s}(\eta)  \partial_\eta$, and the 
differential surface element $dS = 2\pi b^2 d\eta/(\varepsilon g_{1,s})$ on the spheroid.
Then, using Fair and Anderson's 
integral expression~[18] for particle velocity based on slip velocity we obtain
\begin{align}
\bm{\mathcal{U}}
 &= 
-{1\over 3{\cal V}} \int_{S} (\hat{n}\cdot {\bm r})\,{\bm v}_\text{slip} \, dS 
=
- \hat{\bf e}_z {\mu_{ph} \over 4\pi b^2 a} 
\int_{S}
 (\hat{n}\cdot {\bm r})
(\hat{\bf e}_z\cdot {\bm \nabla}_{\!s}{\cal L}\{{\cal D}^{-1}\Gamma\})
 \, dS 
\nonumber \\
 &=  
- \hat{\bf e}_z {\mu_{ph} \over 4\pi b^2 a} 
\int_{-1}^1
\left[a \varepsilon g_{1,s}(\eta)\right]
\left[a^{-1} g_{2,s}^2(\eta)  \partial_\eta({\cal L}\{{\cal D}^{-1}\Gamma\})\right]
{2\pi b^2 d\eta \over \varepsilon g_{1,s}}
=  
- \hat{\bf e}_z {\mu_{ph} \over 2 a} 
\int_{-1}^1
g_{2,s}^2(\eta)  \partial_\eta({\cal L}\{{\cal D}^{-1}\Gamma\})\,
 d\eta
 \label{eq:velintermediate}
 \end{align}
If we can write $g_{1,s}\partial_\eta g_{2}^2$ in terms of 
the normal gradient $g_{1,s}\partial_\xi$ of a harmonic field, 
we can use the fact that $\int_s dS\, \alpha\, {\cal L}\beta =
\int_s dS\, \beta\, {\cal L}\alpha$ for any surface fields 
$\alpha$, $\beta$ in the domain of ${\cal L}$.
So,
\begin{align}
g_{1,s} \frac{\partial g_2^2}{\partial \eta} 
&= 
g_{1,s}{1 \over \ecc^2} \frac{\partial}{\partial \eta}
\left({1 -\eta^2 \over \xi^2 - \eta^2} \right)
= 
g_{1,s}{1 \over \ecc^2}
\frac{-2\eta (\xi^2-1)}{(\xi^2-\eta^2)^2} 
=
g_{1,s}{ \eta (\xi^2-1) \over \xi \ecc^2}
\frac{-2\xi}{(\xi^2-\eta^2)^2} 
=
g_{1,s}{ \eta (\xi^2-1) \over \xi \ecc^2}
\partial_\xi \left[\frac{1}{\xi^2-\eta^2} \right]
\nonumber \\
&=
g_{1,s}{ \eta (\xi^2-1) \over \xi \ecc^2}
{1 \over 2\eta}
\partial_\xi \left[\frac{1}{\xi-\eta} - \frac{1}{\xi+\eta}  \right]
\stackrel{S}{=}
g_{1,s}{ (\ecc^{-2}-1) \over 2 \ecc}
\partial_\xi \left[\frac{1}{\xi-\eta} - \frac{1}{\xi+\eta}  \right]
\nonumber \\
&\stackrel{S}{=}
g_{1,s}{ 1- \ecc^2 \over 2 \ecc^3}
\partial_\xi \left[\frac{1}{\xi-\eta} - \frac{1}{\xi+\eta}  \right]
\stackrel{S}{=}
{ 1- \ecc^2 \over 2 \ecc^3} \,
a \hat{n}\cdot{\bm \nabla} \left[\frac{1}{\xi-\eta} - \frac{1}{\xi+\eta}  \right]
\label{eq:transformed-mystery}
\end{align}
where `$S$' above the equals sign means the equality holds only at the particle surface. 
Now, we need to show that $1/(\xi-\eta)$  and $1/(\xi+\eta)$ are harmonic;
then the function in parentheses in the final expression of (\ref{eq:transformed-mystery}) will be also.
We just make a straightforward check.
In spheroidal coordinates, the Laplacian acting on an axisymmetric field is
\begin{equation}
\nabla^2 =
\frac{1}{\ecc^2(\xi^2-\eta^2)} 
\left[ 
    \frac{\partial}{\partial \xi} (\xi^2-1) \frac{\partial}{\partial \xi}
    + \, \frac{\partial}{\partial \eta} (1-\eta^2) \frac{\partial}{\partial \eta} 
\right].
\end{equation}
Hence, for differentiation with respect to $\eta$ we have 
\begin{align}
\frac{\partial}{\partial \eta} (1-\eta^2) \frac{\partial}{\partial \eta} \frac{1}{\xi-\eta}
&= \left[(1-\eta^2) \frac{\partial^2}{\partial \eta^2} 
-2\eta \frac{\partial}{\partial \eta} \right]
\frac{1}{\xi-\eta}
=
\frac{2(1-\eta^2)}{(\xi-\eta)^3}
- \frac{2\eta}{(\xi-\eta)^2}
=
2\frac{1-\eta\xi}{(\xi-\eta)^3}.
\label{eq:half-Laplacian}
\end{align}
Note that if we swap $\xi$ and $\eta$ in this final expression, the sign is changed.
On the other hand,
\begin{equation}
\frac{\partial}{\partial \xi} (\xi^2-1) \frac{\partial}{\partial \xi} \frac{1}{\xi-\eta}
=
\frac{\partial}{\partial \xi} (1-\xi^2) \frac{\partial}{\partial \xi} \frac{1}{\eta-\xi},
\end{equation}
and this is exactly the left-hand side of (\ref{eq:half-Laplacian}) with $\xi$ and $\eta$
swapped. Thus, this produces something that exactly cancels the term we already
computed. And, therefore, $\nabla^2(\xi-\eta)^{-1} = 0$. Since reflection across the 
$x$-$y$ plane is a symmetry of the Laplacian, it follows that $1/(\xi+\eta)$ and 
the function in parentheses in (\ref{eq:transformed-mystery}) are harmonic as well.

Now, we can evaluate the integral in  Eq.~(\ref{eq:velintermediate}) using $g_{2,s}(\pm 1)=0$ and integrating by parts,
\begin{align}
\int_{-1}^1 d\eta\, g_{2,s}^2 \partial_\eta({\cal L}\{{\cal D}^{-1}\Gamma\})
& =
g_{2,s}^2{\cal L}\{{\cal D}^{-1}\Gamma\}\Big|_{-1}^1 
-\int_{-1}^1 \frac{\partial g_{2,s}^2}{\partial \eta}{\cal L}\{{\cal D}^{-1}\Gamma\} \, d\eta
=
-{\varepsilon \over 2\pi b^2} 
\int_{-1}^1 g_{1,s} \frac{\partial g_{2,s}^2}{\partial \eta}{\cal L}\{{\cal D}^{-1}\Gamma\} \, dS
\nonumber \\
&=
-{\varepsilon \over 2\pi b^2} 
\int_{-1}^1 
({\cal D}^{-1}\Gamma)\, {\cal L}\{g_{1,s} \partial_\eta g_{2,s}^2\} \, dS
=
-\int_{-1}^1 ({\cal D}^{-1}\Gamma)\, {1 \over g_{1,s} } {\cal L}\{g_{1,s} \partial_\eta g_{2,s}^2\} \, d\eta
\nonumber \\
&=
- { 1- \ecc^2 \over 2 \ecc^3} \int_{-1}^1 ({\cal D}^{-1}\Gamma)\, {1 \over g_{1,s} }
 {\cal L}\left\{a \hat{n}\cdot{\bm \nabla} \left[\frac{1}{\xi-\eta} - \frac{1}{\xi+\eta}  \right]\right\} \, d\eta
 \nonumber \\
&=
a { 1- \ecc^2 \over 2 \ecc^3} \int_{-1}^1 ({\cal D}^{-1}\Gamma)\, {1 \over g_{1,s} }
 \left[\frac{1}{\xi_s-\eta} - \frac{1}{\xi_s+\eta}  \right] \, d\eta
 \nonumber \\
&=
a \int_{-1}^1 ({\cal D}^{-1}\Gamma)\, 
 (\xi_s^2-1) \frac{\eta}{\xi_s^2-\eta^2}\sqrt{\xi_s^2 - \eta^2 \over \xi_s^2 - 1} \, d\eta
 \nonumber \\
&=
a \int_{-1}^1 ({\cal D}^{-1}\Gamma)\, \eta \sqrt{\xi_s^2 - 1 \over \xi_s^2 - \eta^2}\, d\eta
\label{eq:Eq10}
\end{align}
Also, defining $\ell = a/b$ we have
\begin{align}
{\xi^2 - 1 \over \xi^2 - \eta^2}
&=
{1- \ecc^2 \over 1 - \ecc^2\eta^2}
=
{1- (1-\ell^{-2}) \over 1 - (1-\ell^{-2})\eta^2}
=
{\ell^{-2} \over 1 - (1-\ell^{-2})\eta^2}
=
{1 \over \ell^2 - (\ell^2-1)\eta^2}
=
{1 \over \eta^2 + (1- \eta^2)\ell^2}
\label{eq:kernelraw}
\end{align}
Plugging 
Eqs.~(\ref{eq:Eq10}) and (\ref{eq:kernelraw}) into (\ref{eq:velintermediate})
 gives
\begin{align}
\bm{\mathcal{U}}
=  
\hat{\bf e}_z {(-\mu_{ph}) \over 2 {\cal D}} 
\int_{-1}^1 {\eta \over \sqrt{\eta^2 + \ell^2(1- \eta^2)}} \, \Gamma d\eta
 \end{align}
Therefore, we obtain the kernel [Eq.~(2) of the manuscript],
\beq
K(\eta,a/b)= \eta [\eta^2 + (a/b)^2(1- \eta^2)]^{-1/2}
\label{eq:kernel}
\eeq

\section{Closed form expressions}
\label{sec:closed-form}

\subsection{Components}
Noting that
\begin{equation}
K(\eta;\ell) = \frac{1}{1-\ell^2} \frac{d}{d\eta} \left[ \eta^2 + \ell^2(1-\eta^2)\right]^{1/2},
\label{eq:K-antiderivative}
\end{equation}
it immediately follows that
\begin{equation}
\int_{\eta_0}^1 K(\eta;\ell)\,d\eta = - \int_{-1}^{\eta_0} K(\eta;\ell)\,d\eta
  =  {1-\sqrt{\eta_0^2 + \ell^2(1- \eta_0^2}) \over 1- \ell^2 }.
\label{eq:K-integrals}
\end{equation}
For evaluating velocities (as in Section \ref{sec:Popescu}), this suffices.
For efficacies, surface areas are also needed.
Denoting the surface area of the region $\{\eta_0 < \eta \le 1\}$ by $S(\eta_0,1)$, we have
\begin{equation}
  \label{eq:areas}
2\frac{S(\eta_0,1)}{S} = 1 - \frac{ {\cal A}(\ell,\eta_0) }{ {\cal A}(\ell,1) },
\end{equation}
where the auxiliary function ${\cal A}$ is given by 
\begin{equation}
\label{eq:A}
{\cal A}(\ell,\eta) :=
\eta \sqrt{[1 -\ell^{-2}] - \eta^2[1 -\ell^{-2}]^2}
 +  \sin^{-1}(\eta\sqrt{ 1-\ell^{-2}}).
\end{equation}

\subsection{Efficacies}

In the standard source/sink configuration, there is a uniform  
$\alpha_+>0$ over the source region of area $S_+ = S(\eta_0,1)$,
and uniform flux $\alpha_- < 0$ over the sink region of area $S_- = S - S_+$.
With total absolute flux $\|\Gamma\|:=\int_S |\Gamma|\,dS = \alpha_+ S_+ - \alpha_- S_- $,
the condition of zero net flux $\int_S  \Gamma \,dS = 0 = \alpha_+ S_+ + \alpha_- S_- = 0$
yields $\alpha_\pm = \pm\|\Gamma\|/(2S_\pm)$. Using Eqs. (\ref{eq:K-integrals}) and (\ref{eq:areas}),
we then find
\begin{align}
{\cal N}^{s/s}(\eta_0,\ell) 
& = 
{S \over \|\Gamma\|}\int_{-1}^1 K(\eta;\ell)\Gamma(\eta)\,d\eta
=
{S \over \|\Gamma\|} \left[ - {\|\Gamma\|\over 2 S_-} \int_{-1}^{\eta_0} K(\eta;\ell)\,d\eta +  {\|\Gamma\|\over 2S_+} \int_{\eta_0}^1 K(\eta;\ell)\,d\eta \right]
\nonumber \\
& =
 S \left({1\over 2S_-}+{1 \over 2S_+}\right) \int_{\eta_0}^1 \!K(\eta;\ell)\,d\eta
 =
 {S^2 \over 2 S_- S_+}  \int_{\eta_0}^1 \!K(\eta;\ell)\,d\eta 
 \nonumber \\
& =
 {2 \over 1 - [{\cal A}(\ell,\eta_0)/{\cal A}(\ell,1)]^2}   
   {1-\sqrt{\eta_0^2 + \ell^2(1- \eta_0^2}) \over 1- \ell^2 }
\end{align}
Similarly for source/inert configuration with active surface $\{\eta \geq \eta_0\}$ 
of area $S_+$ with uniform flux $\|\Gamma\|/S_+$
we obtain
\begin{align}
{\cal N}^{s/i}(\eta_0,\ell) 
& = 
{S \over \|\Gamma\|}\int_{-1}^1 K(\eta;\ell)\Gamma(\eta)\,d\eta
=
{S \over \|\Gamma\|}  \int_{\eta_0}^1 K(\eta;\ell)  {\|\Gamma\|\over S_+} \,d\eta
=
{S \over S_+}  \int_{\eta_0}^1 K(\eta;\ell)   \,d\eta
 \nonumber \\
& =
 {2 \over 1 - {\cal A}(\ell,\eta_0)/{\cal A}(\ell,1)}   
   {1-\sqrt{\eta_0^2 + \ell^2(1- \eta_0^2}) \over 1- \ell^2}
 \end{align}

\section{Self-diffusiophoresis}
\label{sec:Popescu}

\begin{figure}[t]
\begin{center}
\includegraphics[width=0.9\textwidth]{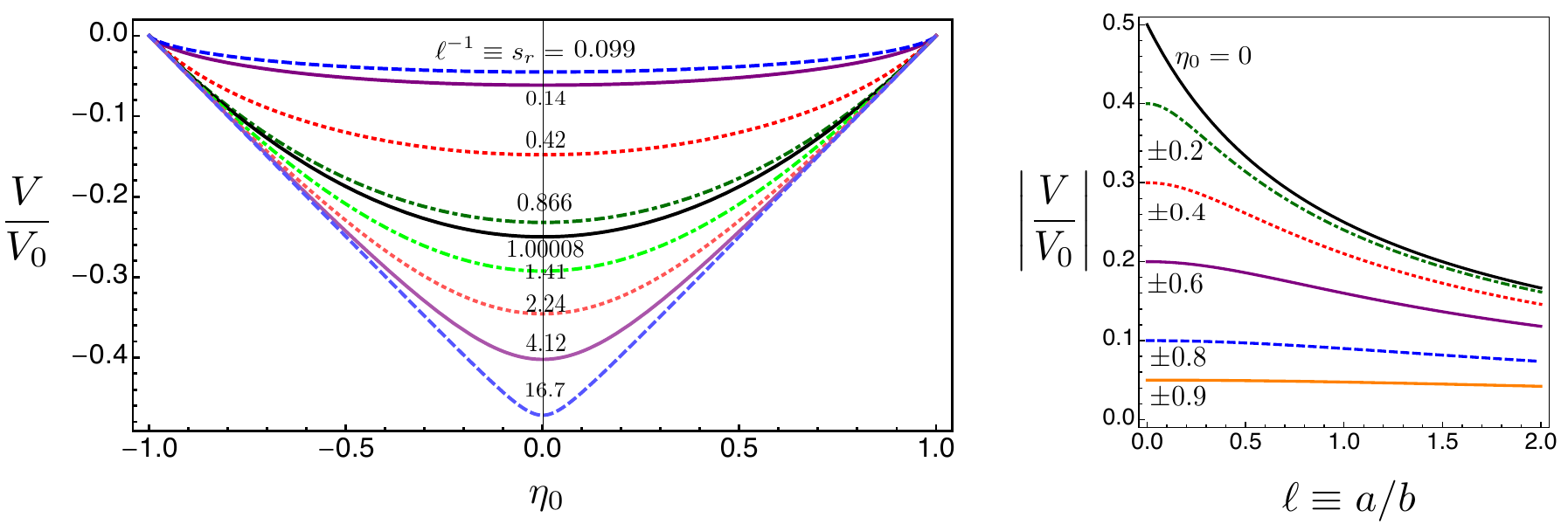}
\end{center}
\vspace{-15pt}
\caption{ 
Left:  plot of Eq.~(\ref{eq:PopescuEquivalence}) corresponding to $s_r \equiv \ell^{-1}$ 
values in Fig.~(4) of~[5]. 
The speed is an even function of $\eta_0$. 
Right: plot of Eq.~(\ref{eq:PopescuEquivalence})  as a function of aspect ratio for 
various values of $\eta_0$. 
The speed is monotonically decreasing with increase in aspect ratio. 
\label{fig:SI-fig}}
\end{figure} 

In this section, we solve Popescu {\em et al.}'s model~[5] of the bipartite 
source/inert self-diffusiophoresis, 
which has uniform flux $\alpha_+$ on the region $\{-1\le \eta \leq \eta_0\}$ and 
zero flux elsewhere. 
Our notations are converted to those of~[5] by the equivalences
$\mu_{ph} \equiv -b$, ${\cal U} \equiv V$, $\alpha_+ \equiv \nu_B \sigma$, 
${\cal D} \equiv D$.
From Eqs.~(\ref{eq:velocity-K-integral}) and (\ref{eq:kernel}) we obtain
the speed of this model as
\begin{align}
V
=  
 {(-\mu_{ph}) \alpha_+ \over 2 {\cal D}} 
\int_{-1}^{\eta_0} {\eta \over \sqrt{\eta^2 + \ell^2(1- \eta^2)}}\,  \, d\eta
=
{b \nu_B \sigma \over 2 D} 
\left[{1- \sqrt{\eta_0^2 + \ell^2(1- \eta_0^2)} \over \ell^2 -1}\right]
=
{V_0 \over 2} 
\left[{1- \sqrt{\eta_0^2 + \ell^2(1- \eta_0^2)} \over \ell^2 -1}\right],
\label{eq:PopescuEquivalence}
\end{align}
where $V_0 = b\nu_B\sigma/D$.
This covers both the prolate and oblate cases which are given in series expansion form in
Eqs.~(32) and (34), respectively, of Ref.~[5].
Limiting behaviors are extracted from this as (the spherical case, $\ell = 1$, is
exact) 
\begin{equation}
  \label{eq:limiting-cases}
V \simeq \begin{cases}
-(V_0/2)(1-|\eta_0|), & \ell\to0^+, \\
 -(V_0/4)(1-\eta_0^2), & \ell = 1, \\
-(V_0/2) \ell^{-1} \sqrt{1-\eta_0^2}, & \ell \gg 1.
\end{cases}
\end{equation}
Equation (\ref{eq:PopescuEquivalence}) and its limiting cases are all even in $\eta_0$, and for a given aspect ratio $\ell$ the maximum speed  $|V|_\text{max}$ always belongs to antisymmetric ($\eta_0=0$) configuration.  Also, for each value of $\eta_0$, the maximum speed is for  extreme discotic case $\ell\to 0^+$ and speed $|V|$ decreases monotonically with increase in aspect ratio $\ell$ as shown in Fig. S1.  \\ \\ \\ 

\noindent
{\bf References}\\
\indent
$[5]$ ~~M. N. Popescu, S. Dietrich, M. Tasinkevych, and J. Ralston, Eur. Phys. J. E {\bf 31}, 351 (2010).\\
\indent
$[7]$ ~~A. Nourhani, V. H. Crespi, and P. E. Lammert, Phys. Rev. E {\bf 91}, 062303 (2015).\\
\indent
$[18]\,$ M. C. Fair and J. L. Anderson, Journal Of Colloid And Interface Science {\bf 127}, 388 (1989).\\